\documentclass[journal=jacsat,manuscript=article]{achemso}

\usepackage[version=3]{mhchem} 

\usepackage{amsmath}
\usepackage{amssymb}
\usepackage{bm}
\usepackage{color}
\usepackage{graphics}
\usepackage{graphicx}
\usepackage{setspace}

\author{Shaozhen Li}
\affiliation[Materials Science Division, Argonne National Laboratory]
{Materials Science Division, Argonne National Laboratory, Argonne, IL 60439 USA}
\alsoaffiliation[School of Physics and Institute for Quantum Materials, Hubei Polytechnic University]
{School of Physics and Institute for Quantum Materials, Hubei Polytechnic University, Huangshi 435003, P. R. China}
\alsoaffiliation[\textbf{S.L., W.Z., and J.D. contributed equally to this work.}]
{ }
\email{shaozhenli@yahoo.com}

\author{Wei Zhang}
\affiliation[Materials Science Division, Argonne National Laboratory]
{Materials Science Division, Argonne National Laboratory, Argonne, IL 60439 USA}
\alsoaffiliation[\textbf{S.L., W.Z., and J.D. contributed equally to this work.}]
{ }
{}

\author{Junjia Ding}
\affiliation[Materials Science Division, Argonne National Laboratory]
{Materials Science Division, Argonne National Laboratory, Argonne, IL 60439 USA}
\alsoaffiliation[\textbf{S.L., W.Z., and J.D. contributed equally to this work.}]
{ }
{}

\author{John E. Pearson}
\affiliation[Unknown University]
{Materials Science Division, Argonne National Laboratory, Argonne, IL 60439 USA}

\author{Valentine Novosad}
\affiliation[Unknown University]
{Materials Science Division, Argonne National Laboratory, Argonne, IL 60439 USA}

\author{Axel Hoffmann}
\affiliation[Unknown University]
{Materials Science Division, Argonne National Laboratory, Argonne, IL 60439 USA}

\title[An \textsf{achemso} demo]
  {Epitaxial patterning of nanometer-thick Y$_\mathrm{3}$Fe$_\mathrm{5}$O$_\mathrm{12}$ films with low magnetic damping}

\abbreviations{IR,NMR,UV}

\begin{document}

\begin{tocentry}

Some journals require a graphical entry for the Table of Contents.
This should be laid out ``print ready'' so that the sizing of the
text is correct.

Inside the \texttt{tocentry} environment, the font used is Helvetica
8\,pt, as required by \emph{Journal of the American Chemical
Society}.

The surrounding frame is 9\,cm by 3.5\,cm, which is the maximum
permitted for  \emph{Journal of the American Chemical Society}
graphical table of content entries. The box will not resize if the
content is too big: instead it will overflow the edge of the box.

This box and the associated title will always be printed on a
separate page at the end of the document.

\end{tocentry}


\clearpage

\begin{abstract}
Magnetic insulators such as yttrium iron garnet, Y$_\mathrm{3}$Fe$_\mathrm{5}$O$_\mathrm{12}$, with extremely low magnetic damping have opened the door for low power spin-orbitronics due to their low energy dissipation and efficient spin current generation and transmission. We demonstrate reliable and efficient epitaxial growth and nanopatterning of Y$_\mathrm{3}$Fe$_\mathrm{5}$O$_\mathrm{12}$ thin-film based nanostructures on insulating Gd$_\mathrm{3}$Ga$_\mathrm{5}$O$_\mathrm{12}$ substrates. In particular, our fabrication process is compatible with conventional sputtering and liftoff, and does not require aggressive ion milling which may be detrimental to the oxide thin films. Structural and magnetic properties indicate good qualities, in particular low magnetic damping of both films and patterned structures. The dynamic magnetic properties of the nanostructures are systematically investigated as a function of the lateral dimension. By comparing to ferromagnetic nanowire structures, a distinct edge mode in addition to the main mode is identified by both experiments and simulations, which also exhbits cross-over with the main mode upon varying the width of the wires. The non-linear evolution of dynamic modes over nanostructural dimensions highlights the important role of size confinement to their material properties in magnetic devices where Y$_\mathrm{3}$Fe$_\mathrm{5}$O$_\mathrm{12}$ nanostructures serve as the key functional component. \\


\end{abstract}


\clearpage

\section{Introduction}

Magnetic insulators such as yttrium iron garnet, Y$_\mathrm{3}$Fe$_\mathrm{5}$O$_\mathrm{12}$ (YIG) with extremely low magnetic damping have been indispensable in almost every aspect of contemporary spin-orbitronics research \cite{ssp_2013,hoffmann_prapplied}. For many decades, the growth of single crystal YIG films has been dominated by liquid phase epitaxy (LPE) \cite{linares_jap}, which yields films in the thickness range from several hundreds of nanometers to millimeters. On the other hand, the exotic magnetic properties revealed recenlty from various YIG-based magnetic heterostructures \cite{kajiwara_nature,heinrich_prl,chumak_ncomm,qu_prl,nakayama_prl} call for urgent need for their nanostructured forms in order to realize practical devices for applications, including spin transfer torque devices \cite{hamadeh_prl,sklenar_prb,jungfleisch_arxiv}, magnetic logic devices \cite{chumak_nphys,ding_apl2012}, auto-oscilators \cite{hamadeh_prl,demidov_nmat}, and skyrmion memories \cite{jiang_science}. However, the conventinal LPE method is incompatible with current industrial top-down nanofabrication technologies. Recently, great advances have been achieved in the growth of high quality YIG films using pulsed laser deposition (PLD) and magnetron sputtering at elevated temperatures \cite{sun_apl,kelly_apl,liu_jap,chang_ieee,wanghl_prb,Kehlberger_prapplied,onbasli_aplm}, yielding nanometer-thick films with low magnetic damping similar to single-crystal YIG bulk materials. For example, a Gilbert damping constant of $\alpha$ = 0.00023 has been achieved by Sun \textit{et al} \cite{sun_apl} and by d'Allivy Kelly \textit{et al} \cite{kelly_apl} using PLD, and $\alpha \sim$ 0.00009 has been achieved by Chang \textit{et al} \cite{chang_ieee} using magnetron sputtering. These demonstrations, more suited for commercial production, are of great technological significance. 

Despite the intensive investigations on YIG continuous films, an efficient yet reliable method for epitaxial patterning of YIG nanostructures is still missing. Microstructured YIG films have been prepared in the past using aggressive ion etching of sputtered YIG films while using resist mask to define the morphological structures \cite{hamadeh_prl,hahn_apl,jungfleisch_jap}. However, the injected, highly energized Ar ions are also detrimental to the films, particularly harmful to oxides with higher stiffness as opposed to metals \cite{wz_review}. Mechanical defects and even cracks, as well as non-trival modifications of the magnetic properties such as saturation magnetization and damping constant have been observed after the ion milling process \cite{jungfleisch_jap}. In this work, we report high-quality growth of YIG films using combined room temperature (RT) magnetron sputtering and \textit{ex-situ} post annealing. In particular, our approach is also compatible with modern nano-lithgraphy techniques, taking advantage of the RT deposition. We demonstrate epitaixal patterned YIG nanostructures using electron beam lithography and liftoff. A Gilbert damping constant comparable to the extended thin-films is achieved even for the patterned structures.

\section{Results and discussion}

The YIG films are deposited on (111)-oriented gadolinium gallium garnet (Gd$_\mathrm{3}$Ga$_\mathrm{5}$O$_\mathrm{12}$, GGG) single crystal substrates at room temperature (RT) from a commercial YIG sputter target. The Ar gas flow, chamber pressure, and sputtering power are kept at 16 sccm, 10 mT, and 75 W, respectively, which is the optimal deposition enviroment for RT growth. \textcolor{black}{It has been shown previously that these films can display different surface morphologies \cite{syvorotka_jmmm} depending on the growth conditions. Here, we used very low sputtering rate and deposited our films from a stoichiometric YIG targets under ultrahigh vacuum. In addition, our sputtering system has an on-axis geometry which does not induce any substrate misorientation.} The as-grown films have a dark gray color implying that the stoichiometry has changed during the RT deposition. The films are subsequently subjected to an \textit{ex-situ} post annealing at 800 $^{\circ}$C for 2 hours in a tube furnance with continuous air flow. The temperature ramping rate is $\sim$ 120 $^{\circ}$C / hour. The key attributes for obtaining a good flim in our process is to ensure an oxygen-rich environment during annealing. Nevertheless, we found that flowing just air at ambient pressure is sufficient enough instead of using pure oxygen. Alternatively, such annealing can also be done \textit{in-situ} after the film growth \cite{liu_jap}; however, much longer annealing time ($>$ 10 hours) is required primarily due to the limitation of the maximum achieveable oxygen partial pressure in the commercial sputtering chamber. The annealed films have a light yellow color (nanometer-thick YIG films). 

Nanostructured YIG films are fabricated by RT deposition onto PMMA/PMGI bilayer resisits defined via electron beam lithography (Raith 150). Since the GGG substrate is highly insulating, we sputtered 5-nm-thick Au after spin coating the bilayer resists to allow efficient electron charge dissipation during the lithography. The Au layer is subsequently removed by gold etcher, followed by wet development of PMMA (MicroChem MIBK) and PMGI (Shipley CD-26), respectively. The PMMA/PMGI bilayer resists can form an undercut cross-section profile which is suitable for magnetron sputtering deposition. After deposition of 40 nm YIG film using the same recipe as before, we remove the resists and surplus materials on top by a resist remover (Shipley 1165). Finally, the nanostructured samples are annealed using the same recipe as for continuous films. Although we choose to use electron beam lithography for our demonstration, the general process is also applicable to other unconventional lithography techniques \cite{wz_all}.     

\begin{figure}
\includegraphics[scale = 0.6]{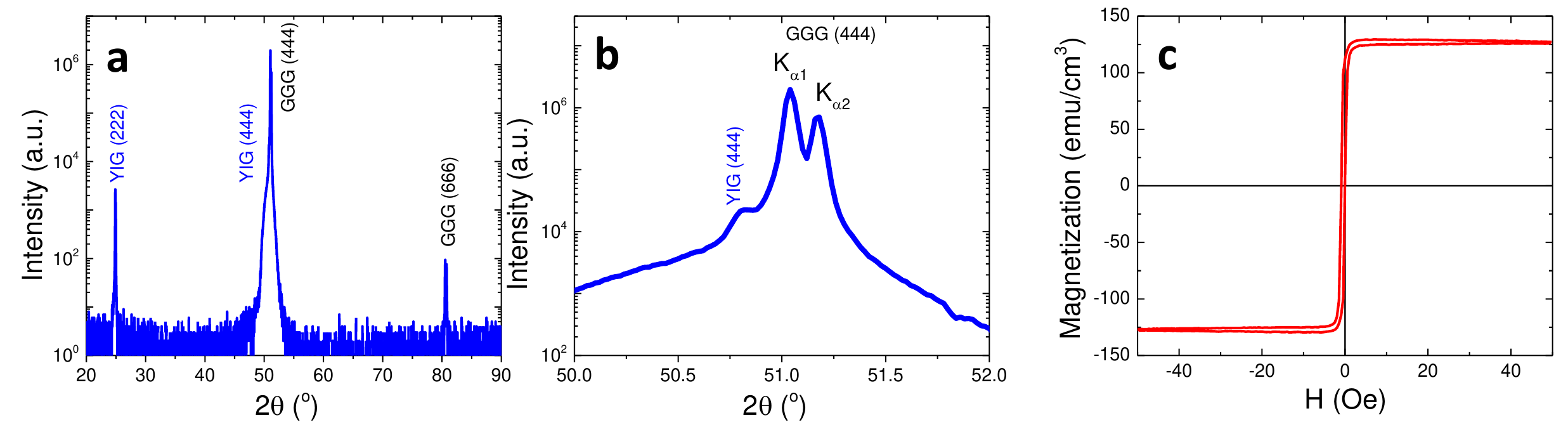}
\caption{(a) X-ray diffraction pattern of a 40 nm annealed YIG film, and (b) the same spectrum in an expanded scale showing the YIG(444) peak. (c) Magnetic hysteresis loop of the same sample measured with an in-plane magnetic field, yielding a saturation magnetization value of $\sim$ 130 emu/cm$^3$. }
\end{figure}

We first characterized our YIG films structurally and magnetically. Figure 1(a) and (b) show the x-ray diffraction pattern our YIG films. The data confirm the (111)-oriented YIG phase in the samples and show no evidence for the existence of any additional phases. Figure 1(c) shows the magnetic hysteresis loop measured by a vibrating sample magnetometry with an in-plane magnetic field. The data indicates very small coercivity, below 1 Oe, and a saturation magnetization, $M_s$ = 130 $\pm$ 20 emu/cm$^3$. This value is in equivalent to a $4 \pi M_s =$ 1633 $\pm$ 251 G, which is only 6$\%$ smaller than the literature value for bulk YIG crystals \cite{ssp_2013}. A more precise value of $4 \pi M_s$ will be determined via dynamic measurements below. 

\begin{figure}
\includegraphics[scale = 0.7]{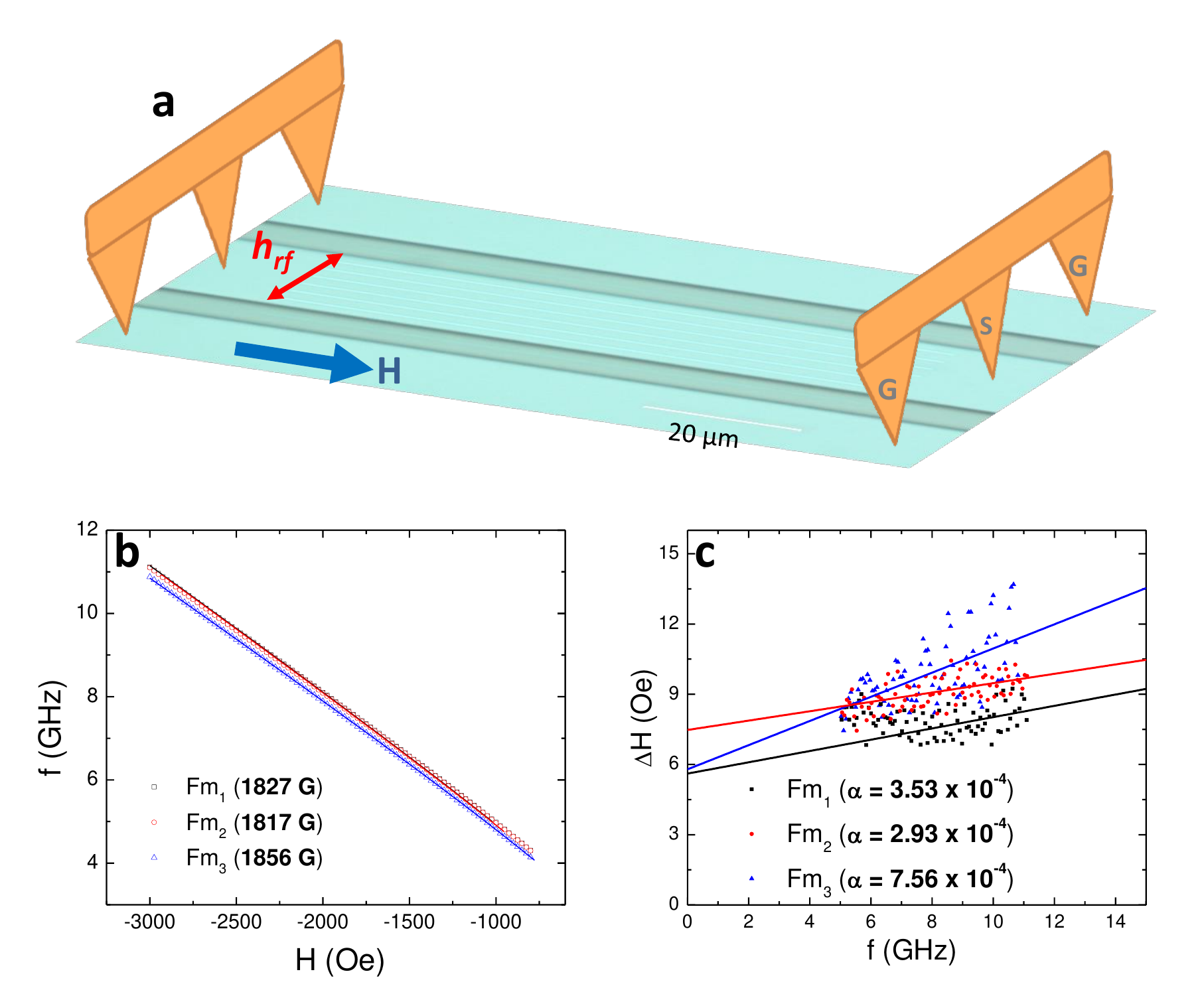}
\caption{\textcolor{black}{(a) VNA-FMR measurement configuration showing the directions of static and dynamic magnetic fields with respect to the samples.} Ferromagnetic resonance data of three YIG large bars (Fm$_{1,2,3}$) showing (b) resonance frequency versus field and corresponding fitting with Kittel equation, and (c) resonance linewidth versus frequency and corresponding linear fitting. }
\end{figure}

Dynamic magnetic properties are investigated by a vector-network-analyzer ferromagnetic-resonance method (VNA-FMR) in a probe station using low-loss rf probes. We make 18 $\mu$m $\times$ 2 mm large YIG bars using the previously described patterning method. Considering that the thickness of the YIG film is only 40 nm, these structures are large enough to be considered as continuous films, which also serve as the reference samples for our YIG nanostructures as will be discussed later. On chip coplanar-waveguides made from Ti(5 nm)/Au(150 nm) are subsequently fabricated on top of the YIG bars by photolithography and liftoff. \textcolor{black}{The measurement configuration is illustrated in Figure 2(a).} We show in Figure 2(b) and (c) the ferromagnetic resonance data of three different YIG samples (Fm$_{1,2,3}$, where Fm denotes `Film'). The resonance field, $H_{FMR}$, versus frequency, \textit{f}, can be fitted by the Kittel equation: 

\begin{equation}
f = |\gamma| \sqrt{H_{FMR} (H_{FMR} + 4 \pi M_s)},
\end{equation}
where $\gamma$ is the gyromagnetic ratio. Our fitting (Fig. 2(b)) yields $4 \pi M_s$ values of 1827, 1817, and 1856 G for Fm$_{1}$, Fm$_{2}$,  and Fm$_{3}$, respectively. On the other hand, the resonance linewidth, $\Delta H$, versus \textit{f} can be linearly fitted with: 

\begin{equation}
\Delta H = \frac{2 \alpha}{|\gamma|} f + \Delta H_0,
\end{equation}
where $\Delta H_0$ denotes the inhomogenous linewidth broadening. Our fitting \cite{kalarickal_jap} [Fig. 2(c)] yields magnetic damping values of 3.53, 2.93, and 7.56 ($\times 10^{-4}$), respectively. The similar $4 \pi M_s$ as well as the low magnetic damping values for these different samples demonstrate the reproducibility of our fabrication process. \textcolor{black}{These values are also comparable to that of similar YIG thin-films fabricated by other approaches \cite{sun_apl,kelly_apl,liu_jap,chang_ieee}.}  

We next move onto the discussions of nanostructured YIG. We make arrays of YIG nanowires (NW), and nanodots (ND) with varying dimensions using the previously described method. \textcolor{black}{Owing to the top-down lithography used here, all nanostructures exhibit quite uniform size and spacing. \cite{adeyeye_jphysd}} As a demonstration, we fabricated NW samples with different widths, denoted as    NW$_{300,450,600,750,1800}$, where the subscript indicates the wire width in nanometer. We also make circular and elliptical dots of similar dimensions. Figure 3 shows the morphology of our nanostructured YIG samples by using scanning electron microscopy after 5 nm Au coating. Clean edges and faithful pattern transfer are achieved thanks to the well-defined undercut resist bilayer. \textcolor{black}{The morphology of these patterned films is also similar to their continuous-film counterparts due to the identical growth and annealing conditions.}

\begin{figure}
\includegraphics[scale = 0.5]{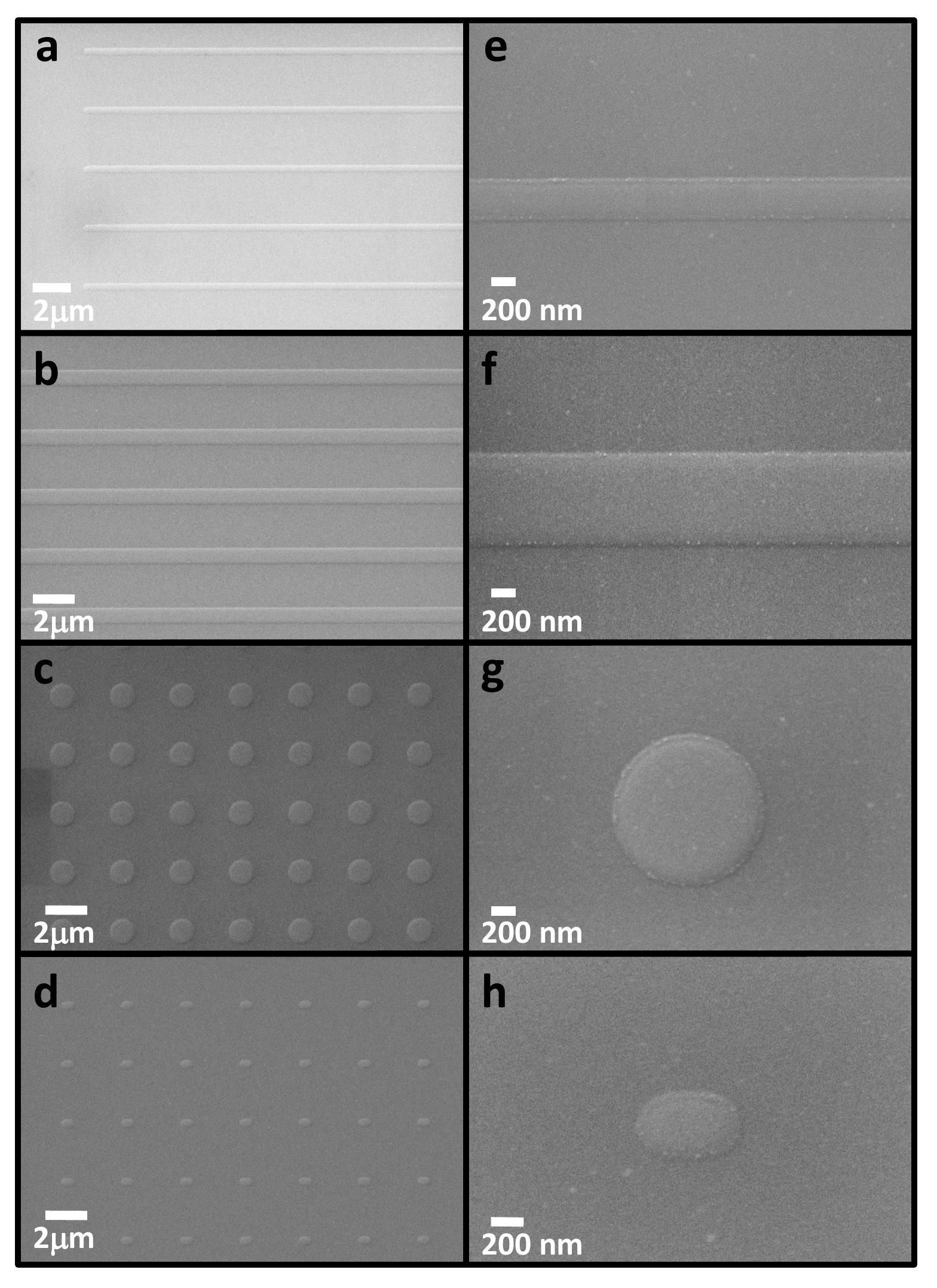}
\caption{Scanning electron microsopy images of epitaxial patterned YIG nanostructures on GGG substrates showing (a) 300-nm-wide wires, (b) 750-nm-wide wires, (c) 600-nm-radius circular dots, (d) 410-nm $\times$ 670-nm elliptical dots, and their corresponding zoomed-in images (e-h).}
\end{figure}

We fabricate on-chip coplanar waveguides on top of each patterned arrays in order to study the dynamic properties of nanostructured YIG. We focus primarily on the NW samples with varying widths so that the effects of geometrical confinement to the magnetization dynamics can be systematically investigated. The gray scale mapping in Figure 4(a) and (b) compare the FMR property of Fm$_1$ and NW$_{300}$. The bright color indicates a low microwave absorption, while the dark color corresponds to a high microwave absorption. The FMR spectrum of the continuous film is symmetric, while the spectrum of NW$_{300}$ exhbits a clear asymmetry with respect to \textit{H} = 0 due to the shape anisotropy originated from the confined edges of the nanostructures. In addition to the main FMR mode, we clearly identify an appreciable edge mode for the NW$_{300}$ sample. Both the main and edge modes are accurately reproduced by our micromagnetic simulations using MuMax Simulator, see Fig. 4(c) and (d). The parameters used for the simulation are: $M_s = 147.77$ emu/cm$^3$, $A_{ex} = 4 \times$ 10$^{-13}$ J/m, $\alpha = 7.561 \times$ 10$^{-4}$, $\gamma = 0.00284$ GHz/Oe. All the values are based on the fitting results of the reference sample. At \textit{H} = -1000 Oe, the simulated edge mode sits at $\sim$ 5.5 GHz [Fig. 4(d)], which agrees very well with the experiment. 

\begin{figure}
\includegraphics[scale = 0.7]{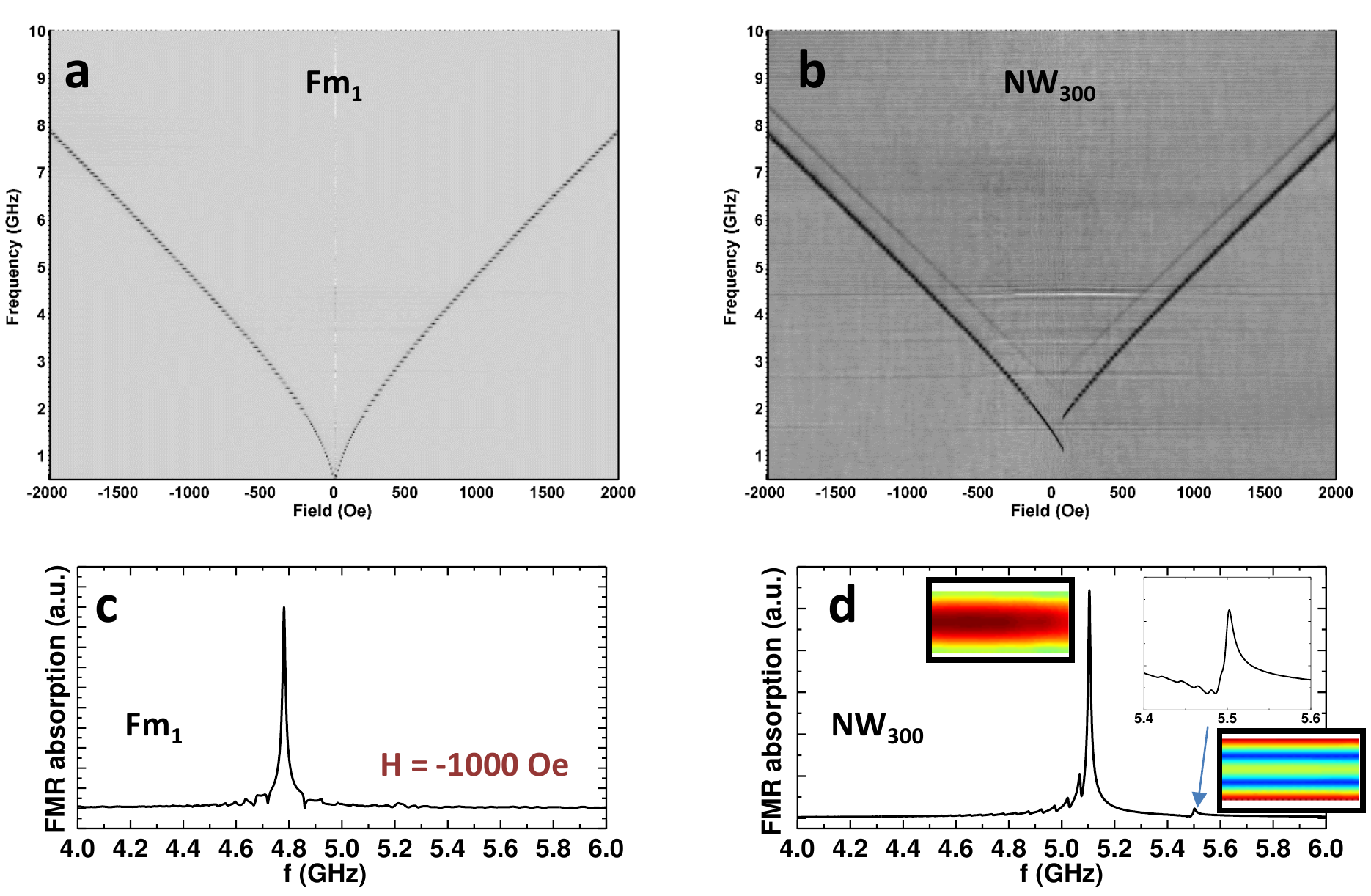}
\caption{FMR 2D-spectrum measured for samples (a) Fm$_1$ and (b) NW$_{300}$, and the corresponding micromagnetic simulations at \textit{H} = -1000 Oe (c-d). Both main and edge modes are identified for NW$_{300}$. Inset pictures of (d) show simulated spatial distribution of magnetization dynamics from the two modes, in which the red color indicates a high spin precession amplitude, and the blue color corresponds to low amplitude.}
\end{figure}

\begin{figure}
\includegraphics[scale = 0.6]{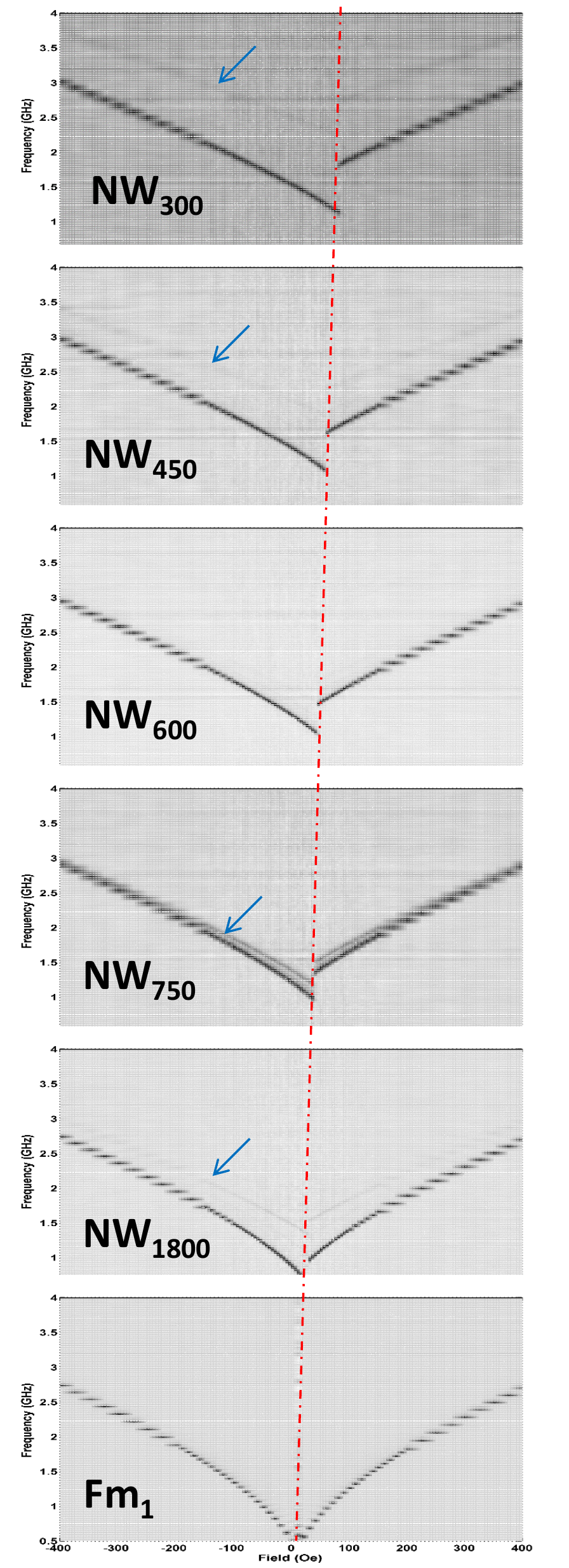}
\caption{FMR 2D-spectrum of Fm$_1$ and NW series of samples, showing evoluation of the coercivity (dashed line) and the magnetization dynamic modes over the width of the wires. Arrows indicate the existence of edge modes. \textcolor{black}{Only the ascending field branch of the hysteresis is shown for simplicity.}}
\end{figure}

Edge modes can be expected for generic nanopatterned magnetic films. However, earlier work has shown that such modes are missing in high damping ferromagnetic metals such as Permalloy (Py, Ni$_{80}$Fe$_{20}$) nanowires with similar structural dimensions \cite{ding_prb2011}. The distinct edge modes of YIG nanostructures observed here is owing to the intrinsic low magnetic damping and weak magnon exchange interactions of YIG, so that they can be well separated from the main FMR mode. 

Figure 5 further demonstrates the evolution of such edge mode with the width of the nanowires. Starting from NW$_{300}$, the edge mode moves closer for NW$_{450}$ as the width increases, and finally merges with the main FMR mode for NW$_{600}$. Notably, this mode starts to reoccur for NW$_{750}$ and NW$_{1800}$ as the width further increases. This cross-over between main and edge modes indicates a higher sensitiviy of the edge mode than the main mode to the change of local demagnetization field since they reside primarily at the edges of the nanowires where the pinning is much stronger as opposed to the center [Fig. 4(d)]. The strong demagnetization field is also evidenced by the fact that even the main mode shows clear evolution with width at \textit{H} = 0 Oe, as shown in Fig. 6(a) and (b). However, when the external field is sufficiently high, no appreciable evolution with width can be observed for the main mode, Fig. 6(b). 

\begin{figure}
\includegraphics[scale = 0.5]{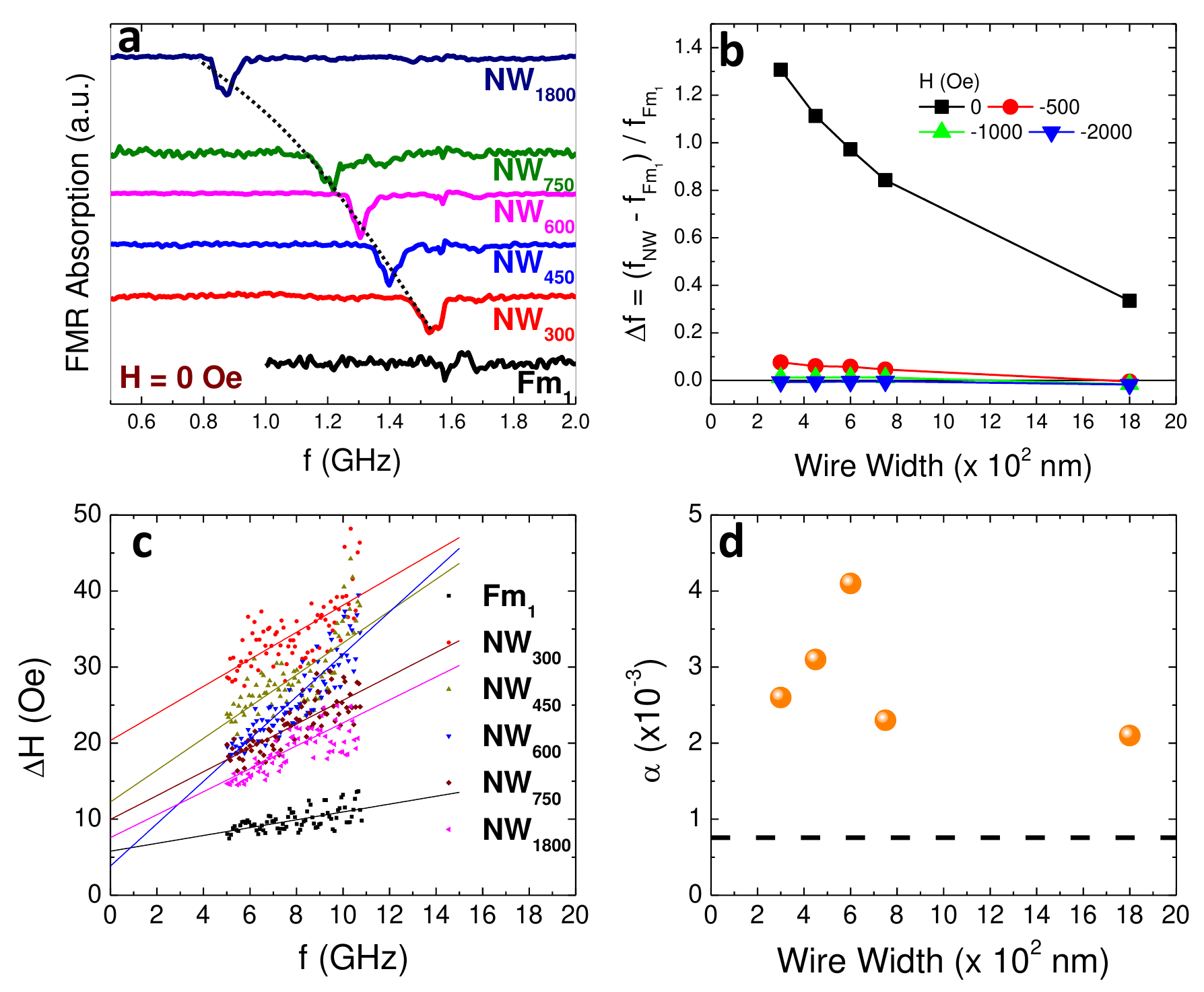}
\caption{(a) FMR 1D-spectrum of Fm$_1$ and NW series samples measured at zero external field showing the evolution of the main FMR mode. (b) Percent change in the resonance frequency of the main FMR mode at different external applied field, \textit{H} = 0, -500, -1000, and -2000 Oe for different width of the nanowire. (c) Resonance linewidth versus frequency of the different samples. (d) Extracted magnetic damping values for nanowires with different width. Dashed line indicates the damping value for the continuous film.}
\end{figure}

The `appear-merge-reoccur' behavior of the edge mode to the main mode (Fig. 5) is an interesting observation, which also finds good agreement with the evolution of the damping constant. We study the frequency dependent resonance linewidth for all our NW samples [Fig. 6(c)] and extract the damping parameters via fitting with Eq.(2), as shown in Fig. 6(d). Starting from NW$_{300}$, the damping constant increases as width increases, and peaks at NW$_{600}$, corresponding to exactly the merge point of the edge and main modes in Fig. 5. This is explained by the fact that when the two modes degenerate, they are coupled due to magnon-magnon scattering leading to efficient energy transfer between the modes \cite{adur_prl}, which as a result has their effective damping increased. The loss of the edge mode at this point is compensated by the large damping enhancement for the main mode. After all,  it is the total linewidth that truly quantifies the losses of magnetic energy regardless of the nature and number of microscopic mechanisms involved. In fact, such degeneracy is something to be avoided for device applications such as for the magnetic auto-oscilators, because it could lead to selflimiting damping and prevent the onset of auto-oscilations. Therefore, our results here highlight the significance and general guidance of proper device engineering for YIG based spintronics.

\section{Conclusions}

Magnetic insulators with low magnetic damping such as Y$_\mathrm{3}$Fe$_\mathrm{5}$O$_\mathrm{12}$ have been widely proposed and investigated as good candidates for pure spin current spintronics concepts. Fundamental studies have reached several milestones in many sub-fields such as spin pumping, spin-Seebeck effect, spin transfer torque, and auto-oscilation. However, successful patterning of nanostructures from such materials for practical device applications has only recently become feasible.    

Here, we have demonstrated reliable and efficient epitaixal growth and nanopatterning of Y$_\mathrm{3}$Fe$_\mathrm{5}$O$_\mathrm{12}$ thin-film based nanostructures on insulating GGG substrates. Strucutral and magnetic properties indicate good qualities, in particular low magnetic damping of both films and patterned structures. We systematically studied the evolution of the dynamic magnetization parameters over the change of the lateral dimensions. A distinct edge mode in addition to the main mode is identified by both experiments and simulations, which also exhbit cross-over with the main mode over changing the width of the wires. The cross-over leads to also a significantly enhanced magnetic damping. The non-linear evolution of dynamic modes over nanostructural dimensions highlights the important role of size confinement to their material properties in magnetic devices where YIG nanostructures serve as the key functional component.

\begin{acknowledgement}

We thank Dr. Jennifer Zheng and Dr. Junjie Zhang for technical help for using the tube furnace. Work at Argonne was supported by the U.S. Department of Energy, Office of Science, Materials Science and Engineering Division. Use of the Center for Nanoscale Materials was supported by the U. S. Department of Energy, Office of Science, Basic Energy Sciences, under Contract No. DE-AC02-06CH11357. S.L. acknowledge the National Natural Science Foundation of China (No. 51302074 and 11374147), the Natural Science Foundation of Hubei Province (No. 2012FFB010), the Creative team of Hubei Polytechnic University of China (Project No. 13xtz05), and the Education Foundation of Hubei Province (D20144402).\\
\\\textit{Conflict of Interest}: The authors declare no competing financial interest.

\end{acknowledgement}




\end{document}